\documentclass[grl]{agu2001}
\usepackage{agu-ps}
\usepackage[dvips,final]{graphicx}
\ifgalley
\setlastpagenum{4}
\fi
\newcommand{\be}{\begin{equation}}
\newcommand{\ee}{\end{equation}}

\newcommand{\eqref}[1]{eq.\,(\ref{#1})}

\newenvironment{eq}{\vspace{2mm}\begin{equation}\begin{array}{l}}{\end{array}
                    \end{equation}\vspace{2mm}}
\authorrunninghead{COSTA AND MACEDONIO }
\titlerunninghead{DEPTH AVERAGED EQUATIONS FOR LAVA FLOWS}

\journalid{}

\articleid{}{}

\paperid{}

\cpright{agu}{2004}

\received{}
\revised{}
\accepted{}
\published{}

\authoraddr{A. Costa,
Osservatorio Vesuviano - INGV,
Via Diocleziano 328, I-80124 Napoli, Italy.
(e-mail: \hbox{costa@}ov.ingv.it)}

\authoraddr{G. Macedonio,
Osservatorio Vesuviano - INGV,
Via Diocleziano 328, I-80124 Napoli, Italy.
(e-mail: \hbox{macedon@}ov.ingv.it)}

\begin{document}
\title{Numerical simulation of lava flows based on depth-averaged
  equations}   

\author{Antonio Costa and Giovanni Macedonio}
\affil{Osservatorio Vesuviano, Istituto Nazionale di Geofisica e
Vulcanologia, Napoli, Italy}
\begin{abstract}
Risks and damages associated with lava flows propagation (for instance 
the most recent Etna eruptions) require a quantitative description of 
this phenomenon and a reliable forecasting of lava flow paths. 
Due to the high complexity of these processes, numerical solution of
the complete conservation equations for real lava flows is often
practically impossible. To overcome the computational difficulties,
simplified models are usually adopted, including 1-D models and
cellular automata. In this work we propose a simplified 2D model
based on the conservation equations for lava thickness and
depth-averaged velocities and temperature which result in first order
partial differential equations.  
The proposed approach represents a good compromise between the full
3-D description and the need to decrease the computational time.
The method was satisfactorily applied to reproduce some analytical
solutions and to simulate a real lava flow event occurred during the
1991-93 Etna eruption.
\end{abstract}
\begin{article}
\section{Introduction}
Depth averaged flow models based on the so-called shallow water
equations (SWE) were firstly introduced by De Saint Venant in 1864 and  
Boussinesq in 1872. Nowdays, applications of the shallow water
equations include a wide range of problems which have important
implications for hazard assessment, from flood simulation
\citep{burgar2002} to tsunamis propagation \citep{heipia2001}. \\
In this paper we propose a generalized set of depth averaged
equations, including an energy equation, to describe lava flow
propagation. We considered lava flow as channelized, i.e. moving lava
has a non-continuous roof and the top represents a free surface open
to the atmosphere.     
\section{Model description}       
The model is based on depth-averaged equations obtained by integrating 
mass, momentum and energy equations over the fluid depth, from the
bottom up to the free surface. 
This approach is valid in the limit $H_*^2/L_*^2\ll 1$ (where $H_*$ is
the undisturbed fluid height and $L_*$ the characteristic wave length
scale in the flow direction). This means that we are dealing with very
long waves or with ``shallow water''. \\
Assuming an incompressible homogeneous fluid and a hydrostatic
pressure distribution, the shallow water equations for an uniform or
gradually varied flow are given by:   
\be
\frac{\partial h}{\partial t}+\frac{\partial (Uh)}{\partial x}+
\frac{\partial (Vh)}{\partial y}= 0
\label{tirante}
\ee
\be
\begin{array}{l}
\displaystyle
\frac{\partial (Uh)}{\partial t}+
\frac{\partial(\beta_{xx}U^2h+gh^2/2)}{\partial x}+ 
\frac{\partial(\beta_{yx}UVh)}{\partial y}= 
\displaystyle
-gh\frac{\partial H}{\partial x}-\gamma U  
\label{velx}
\end{array}
\ee
\be
\begin{array}{l}
\displaystyle
\frac{\partial (Vh)}{\partial t}+
\frac{\partial(\beta_{xy}UVh)}{\partial x}+ 
\frac{\partial(\beta_{yy}V^2h+gh^2/2)}{\partial y}= 
\displaystyle
-gh\frac{\partial H}{\partial y}-\gamma V 
\label{vely}
\end{array}
\ee
where $h$ is the fluid depth measured from the altitude of the terrain 
surface $H$ (bed),  $(U,V)=1/h\int_H^{H+h}{\bf u}(x,y,z)dz$  are the
depth-averaged fluid velocity components, $\beta_{ij}$ are correction
factors (in the range 0.5-1.5) and $\gamma$ is a dimensionless
friction coefficient depending on the fluid rheology  and on the
properties of both flow and bed. The gradients 
$\partial H/\partial x_i$ indicate the channel bottom slopes in both
directions $x$ and $y$ ($x_i=x,y$). The terms on the right sides
represent the so-called source terms. \\ 
In the case of lava, the viscosity is strongly temperature
dependent. For this reason, besides the equations (\ref{tirante}),
(\ref{velx})  and (\ref{vely}), it is necessary to solve the equation   
for the energy conservation. From a computational point of view,
the temperature equation is similar to the pollutant transport
equation \citep{monben99,lev2002}.     
We propose the following heuristic equation for the depth-averaged
temperature $T(x,y)=1/h\int_H^{H+h} T(x,y,z)dz$:  
\begin{eq}
\displaystyle\frac{\partial (Th)}{\partial t}+ \frac{\partial
(\beta_{Tx}UTh)}{\partial x}+ \frac{\partial (\beta_{Ty}VTh)}{\partial
y}= -{\mathcal E}(T^4-T_{env}^4)+\\
\vspace{0.2cm}
-{\mathcal W}(T-T_{env}) 
-{\mathcal H}(T-T_c) +{\mathcal K}(U^2+V^2)\exp{[-b(T-T_{r})]} 
\label{temp} 
\end{eq}
where $T_{c}$ and $T_{env}$ are the temperatures of the lava-ground 
interface and of the external environment respectively, and
$\beta_{Ti}$, ${\mathcal E}$, ${\mathcal W}$, ${\mathcal H}$ and
${\mathcal K}$  are a set of semi-empirical parameters.
Terms on the right side of the equation\,(\ref{temp}) represent the 
radiative, convective and conductive exchanges respectively, while the
last term is due to the viscous heating.     
Moreover, a simple exponential relationship between magma viscosity
and temperature was assumed \citep{cosmac2002}:   
\be
\mu=\mu_{r} \exp{[-b(T-T_{r})]}
\label{visc}
\ee
where $b$ is an appropriate rheological parameter and $\mu_{r}$ is
the  viscosity value at the reference temperature $T_{r}$ (for
instance,  $T_{r}=T_0$ with $T_0$ equal to the emission temperature at
the vent). For the description of a thermal balance in lava flows,
similar to the equation\,(\ref{temp}) see \citet{kessel98}. We do not
explicitly accounted for crystallization and crystallinity-dependence of 
the viscosity, but they are implicitly considered in the determination
of the rheological parameters in (\ref{visc}). 
%
%
Concerning the coefficient $\gamma$ which appears in the
equations\,(\ref{velx}) and (\ref{vely}), we propose a relationship
similar to that used in the viscous regime
\citep{gerper2001,fersal2004}: $\gamma=\kappa_*/[1+\kappa_*
  h/(3\nu_r)]$, where $\kappa_*$ is the Navier friction coefficient,
$\nu_r=\mu_r/\rho$ and $\rho =\mbox{fluid density}$. This relationship 
permits in principle to consider different and general wall friction
conditions and, for instance, the possibility to include viscous
heating effects on lava flow velocity \citep{cosmac2003} by choosing
the appropriate $\kappa_*$ parameterization. By considering the
viscosity dependence on temperature(\ref{visc}) and, for simplicity,
the limit  ${\kappa_* h/(3\nu_r)~\gg~1}$, we obtain:    
\be
\gamma = \frac{3\nu_r }{h} \exp{[-b(T-T_{r})]}
\label{reol}
\ee 
In the following, we estimate the other parameters introduced in
(\ref{temp}) evaluating the corresponding terms of the complete
averaged energy equation. The heat transfer coefficient ${\mathcal H}$
is roughly estimated from the term 
$\kappa\int_H^{H+h}\nabla^2 T(x,y,z)dz$:    
\be
{\mathcal H}\approx n\kappa/h
\label{cond}
\ee
where $\kappa=k/(\rho c_p)$ is the thermal diffusivity ($k$ is the
thermal conductivity and $c_p$ the specific heat) and we approximated
the characteristic thermal boundary layer length as a fraction of the
total thickness: $\delta_T\approx h/n$ where $n$ depends on the
temperature profile ($n\sim 4\div \sqrt{\nu/\kappa}$). \\
According to \citet{piebal86}'s study, for the radiative term, we
assumed:   
\be
{\mathcal E}\approx \epsilon\sigma f/(\rho c_p)
\label{rad}
\ee
where $\epsilon$ is the emissivity, $\sigma$ the Stephan-Boltzmann
constant ($\sigma=5.67\cdot 10^{-8}$Wm$^{-2}$K$^{-4}$) and  $f$ is the
fractional area of the exposed inner core \citep{cribal90}. For
simplicity, in this version of the model we assumed $f$ as a
constant. In real lava flows $f$ may change with time and space
$f=f({\bf x},t)$  and, in  principle, it can be estimated from field
measurements or remote sensing. Further studies should investigate the
sensivity of the model with the temporal and spatial changes of this
quantity.\\   
For the convective term, we adopted \citep{kessel98}:
\be
{\mathcal W}\approx \lambda f/(\rho c_p)
\label{conv}
\ee 
where $\lambda$ is the atmospheric heat transfer coefficient.\\
Finally, for the viscous heating term, we approximate the order of 
magnitude of the quantity   
$\Phi=1/(\rho c_p)\int_H^{H+h}\mu(\partial v/\partial z)^2dz$  
as $\mu_r e^{-b(T-T_{r})} (U^2+V^2) m/h$, 
where we approximated the characteristic velocity boundary layer as 
$\delta_v \approx h/m$; hence:   
\be
{\mathcal K}\approx m\mu_r/(\rho c_p h)
\label{vh}
\ee
where in the case of a parabolic velocity profile $m=12$
\citep{shapea74}. \\
By using the approximations and parameterizations described above, we 
obtain the final system of equations we solve by means of the
numerical method described in the Section\,\ref{numeric}. 
\section{The numerical method}
\label{numeric}
The numerical solution of the equations (\ref{tirante}), (\ref{velx}),
(\ref{vely}) and (\ref{temp}), was achieved by using an algorithm
based on the software package CLAWPACK (available on the web at 
{\tt http://www.amath.wa\-shington.edu/\~{}rlj/claw\-pack.html}). 
CLAWPACK is a public domain software package designed to compute 
numerical solutions to hyperbolic partial differential equations using  
a wave propagation approach \citep{lev2002}.  \\   
The CLAWPACK routines were generalized in order to treat the viscous
friction source term and to solve the energy
equation~(\ref{temp}). The modelling of lava flow over an initially
dry downstream region (dry bed problem) was approached following the
method described in \citet{monben99}.     
All the source terms in the governing equations were treated using a  
Godunov splitting method and, since as a simple explicit
discretization leads to  numerical instabilities
\citep[e.g.\ ][]{amb99,monben99}, all terms were discretized using a
semi-implicit scheme. For instance, the source term in the equation
(\ref{velx}) was discretized as below: 
$$\frac{q_{n+1}-q_n}{\Delta t} = -gh_n\frac{\partial H}{\partial x} - 
\frac{3\nu_{r}q_{n+1}}{h_n^2} e^{-b(T_n-T_{r})} $$
where pedice $n$ indicates the quantities at the time $t_n$, and
$q_n=U_nh_n$. The other source terms were discretized by using a
similar approach. 
\par
\begin{figure}
\hspace{-0.2cm}
\includegraphics[angle=0,width=\hsize]{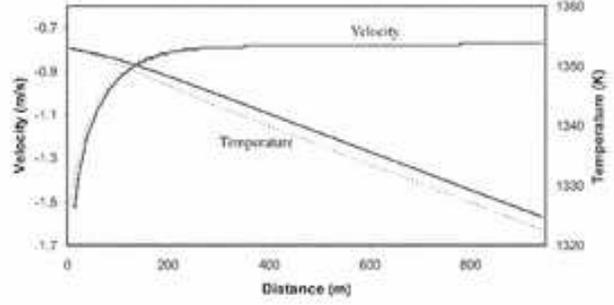}
\caption{Longitudinal profiles of the channel center velocity and
  temperature, at $t=2500$~s. Dashed and continuous lines 
  indicate analytical and numerical results,
  respectively. Channel dimensions: 50~m wide, 1000~m long;
  Slope:~0.1, $T_{env}=0$~K; $T_0=1353$~K, Flow rate: 
  $Q=12.5$~m$^3$/s; $\Delta x = \Delta y = 5$~m. }   
\label{test}
\end{figure}

\begin{figure}
\hspace{-0.25cm}
\includegraphics[angle=0,width=\hsize]{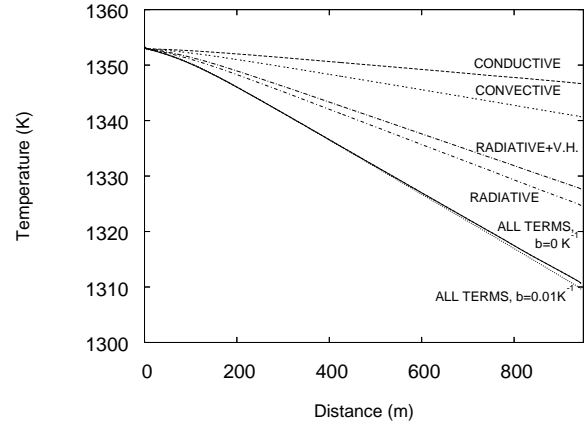}
\caption{Longitudinal temperature profiles at $t=2500$~s obtained 
  using the same parameters of Figure\,\ref{test} and
  {$T_{env}=300$~K}, {$T_c=1173$~K}, {$f=0.5$}, {$\epsilon=0.8$},
  {$n=4$}, {$m=12$}. {V.H. = Viscous  Heating}.}      
\label{allterms}
\end{figure}

\begin{figure}
\hspace{-0.25cm}
\includegraphics[angle=0,width=\hsize]{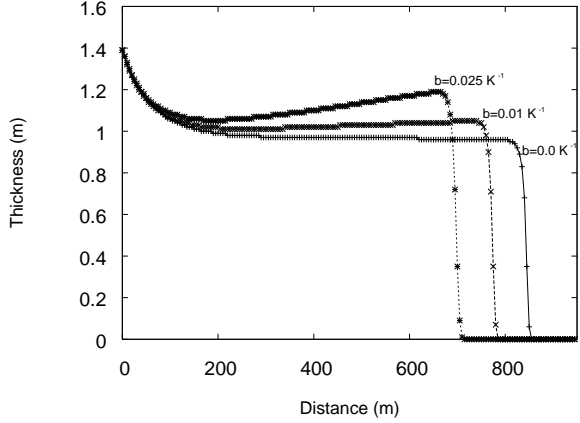}
\caption{Longitudinal thickness profiles at $t=1200$~s obtained
  considering the same parameters reported in
  Figure\,\ref{allterms}.}   
\label{spessori}
\end{figure}

 
Before the application, the algorithm was tested by simulating some
cases for which analytical solutions are known. In fact, considering
the flow of a quasi-unconfined layer of viscous liquid on an inclined
plane, with the energy and the momentum equations decoupled (i.e. with
$b=0$~K$^{-1}$) and in the steady state limit, the equations
(\ref{tirante}), (\ref{velx}), (\ref{vely}) and (\ref{temp}) admit
the following analytical relationships \citep{kessel98,piebal86}:  
\be
q_2 = - q_1^3g\sin\alpha/(3\nu_r)  \quad
q_3 = q_1[T_0^{-3} + 3{\mathcal E}(y-y_0)/q_2]^{-1/3} 
\label{analitico}
\ee
where $q_1=h$, $q_2=hV$, $q_3=hT$, $\alpha$ is the channel slope and
$(y-y_0)$ represents the distance from the vent. Figure\,\ref{test}
shows the comparison between the analytical and numerical
relationships.    
Simulation results have shown a good agreement with an error less than 
1\% for the conservative variables $h$, $hV$ and $hT$ and, within a
few \% for the non-conservative variable $V$ and $T$. 
Moreover, in order to estimate the importance of each term on the
right side of the equation\,(\ref{temp}), we considered the same
geometry of the simple slope flow as above and the typical values
reported in the caption of the Figure\,\ref{allterms}. Results, 
plotted in the Figure\,\ref{allterms}, show that radiative cooling is 
the main heat loss mechanism, while conductive and atmospheric
convective cooling is less important but, for the parameter values
used here, conductive loss is comparable with convection
cooling. Viscous heating effect can be neglected in terms
of mean lava temperature (in the simulated case it produces a increase
of a few $^o$C for a distance of 1~km), although, in certain
conditions, it could be more important and determinant in the
choosing the appropriate wall conditions and exchange coefficients for 
both momentum and energy \citep{cosmac2003}.     
About effects of the coupling between momentum and energy equations,
we can see a non-zero $b$ is important to determine the longitudinal
variation of the lava flow thickness (see Figure\,\ref{spessori}),
although it increases slightly the cooling beyond certain distances.
Figure\,\ref{spessori} shows as the velocity decrease due to 
the longitudinal viscosity increase is able to cause a longitudinal
rise of the lava thickness because of the viscosity temperature
dependence.   
\section{Application to Etna lava flows}
\label{application}
\begin{figure}
\hspace{-0.45cm}
\includegraphics[angle=-90,width=\hsize]{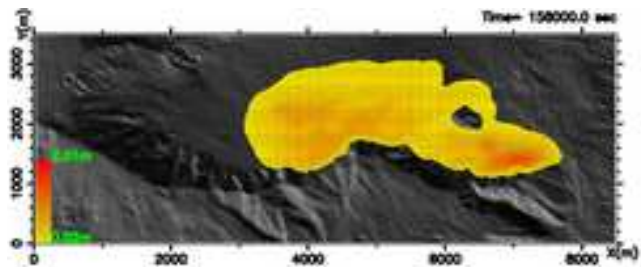}
\caption{Simulated lava thickness of the $3^{rd}$ and $4^{th}$ January
  Etna lava flow.}       
\label{results}
\end{figure}

In this section, as an application, we reported simulation results of
the initial phases of the 1991-1993 Etna eruption for which some field
data for input and comparison are available \citep{calcol94}. In
particular we simulated the second phase occurred from the $3^{rd}$ up
to the $10^{th}$ January 1992. 
In order to estimate previously introduced semi-empirical parameters,
we considered the typical magma parameters reported in
Table~\ref{tab:parametri} partially derived from data of
\citet{calcol94}. We assumed as representative an effective viscosity
of 10$^3$ Pa$\cdot$s at an estimated vent temperature of about
1353~K and $b\approx 0.02$~K$^{-1}$ that, for a cooling of about
100~K, reproduces the observed viscosities of the order of 10$^4$
Pa$\cdot$s \citep{calcol94}. Other parameters were chosen within  
typical ranges: $f=0.1$ (between 0.01 and 1 \citep{kessel98}), and
$\epsilon=0.8$ (between 0.6 and 0.9 \citep{ner98}). $T_c$ is set
higher than its typical values since, for numerical reasons, we need 
to limit the maximum viscosity value.  \\
The parameters reported in Table~\ref{tab:parametri} give the
following typical values:    
\be
\begin{array}{l}
\displaystyle
{\mathcal H}\sim 3/h \times 10^{-6} \, \mbox{m s}^{-1}\\ 
{\mathcal E}\approx  1.5 \times 10^{-15} \,
\mbox{m s}^{-1}\mbox{K}^{-3}\\    
{\mathcal W} \approx 2\times 10^{-6} \, \mbox{m s}^{-1}\\ 
{\mathcal K}\sim 4/h \times 10^{-3} \, \mbox{m s}^{-1}\mbox{K}\\  
\end{array}
\label{eq:numeriadimensionali}
\ee
where, for our aim in this application, we set $T_{env}= 300$~K,
$n=4$, $m=12$ and $\beta_{ij}=1$. \\   
\begin{table}[htb!]
\caption{Parameters characteristic of Etna lava.} 
\vspace{0.2cm}
\renewcommand{\tabcolsep}{2pc} 
\renewcommand{\arraystretch}{1.2} 
\begin{tabular}{ccc}
\hline
$\rho$  & 2500 & kg/m$^3$ \\
$b$     & 0.02 & K$^{-1}$ \\
$c_p$   & 1200 & J\,kg$^{-1}$K$^{-1}$ \\
$k$     & 2.0  & W\,m$^{-1}$K$^{-1}$ \\
$\lambda$ & 70 & Wm$^{-2}$K$^{-1}$\\
$T_c$   & 1253 & K \\
$T_{0}$    & 1353 & K \\
$\mu_r=\mu(T_{0})$ & $10^3$ & Pa\,s\\
\hline
\end{tabular}
\label{tab:parametri}
\end{table}
As topographic basis, we used the digital data files of the Etna
maps with a 1:10000 scale available at the Osservatorio Vesuviano-INGV
web site at {\tt http://venus.ov.ingv.it} (the used spatial grid
resolution was $\Delta~x~=~\Delta~y~=~25$~m).  
For the second phase, we considered an ephemeral vent sited in Piano
del Trifoglietto at the UTM coordinates (503795; 4174843). Finally,
for the period 3-10 January 1992, we considered a constant average
lava flow rate of 16 m$^3$/s (ranging from 8 to 25 m$^3$/s)
\citep{calcol94,barcar93}. \\
The first phase of the eruption corresponded with the initial
spreading of the lava flows on Piano del Trifoglietto. On the $3^{rd}$ 
January 1992 a new lava flow that overlapped the older lava lows,
became an independent branch. By the evening it covered more than
1~km. The day after the front reached Mt. Calanna. One branch
continued to move to the south of Mt. Calanna and one branch turned to 
the north then to the east (see Figure\,3 of
\citet{calcol94}). Because of a significantly decrease of lava supply,
the southern lava flow stopped in Val Calanna. On January $7^{th}$ the 
northern lava lobe touched the southern one and then merged
\citep{calcol94}.\\ 
In Figure\,\ref{results} the simulated lava flow at the end of the
second phase is shown. The model is able to reproduce
semi-quantitatively the behaviour of the real lava flow and the order 
of magnitude of the quantities involved such as thickness, temperature  
and the time of front propagation of the lava flow. 
Although we introduced different simplifications and we considered an  
arduous case encompassing both a large viscous friction term and
complex rough topography, the simulation and real lava flows show 
strikingly similar dynamics and thermal pattern
evolution. Nevertheless the model presented in this paper remains an
initial model of lava flow emplacement using SWE. Future improvements
are expected by refining the the computational performance of the
model and the formulation of the parameters.  
%
\section{Limitations}
This methodology is based on vertical averages and therefore it cannot
be rigorously valid for every conceivable application. We stress that
the model is based on the basic assumptions of (1) small vertical scale
relative to horizontal ($H_*^2/L_*^2\ll 1$), (2) homogeneous
incompressible fluid, (3) hydrostatic pressure distribution, (4) slow
vertical variations. \\  
Concerning the computational method, the principal limit is related to
the numerical treatment we used here for the source terms arising
from topography and viscous friction. In particular since the actual
topographies may contain abrupt variations, the slope term that
appears in the equations\,(\ref{velx}) and (\ref{vely}) can become
infinite in correspondence of discontinuities leading to numerical
oscillations, diffusion, smearing and non-physical solutions
\citep{lev98,alcben2001,chiler2004}. Also the friction term must
be carefully treated. In fact, if the characteristic time of the
source term is much smaller than the characteristic time of the
convective part of the equations, the problem is said to be stiff and
the classical splitting method may provide erroneous physical
solutions on coarse meshes \citep{chiler2004}. 
To avoid these problems a trivial solution is using a very small
time step, which results in long computational times.
%
%
In the next version of the model, this limit could be overcome by
applying directly a method based on the solution of the inhomogeneous 
Riemann problem with source term instead of applying the splitting
method \citep{chiler2004,geo2004}.      
\section{Conclusion}
A new general computational model for lava flow propagation based on
the solution of depth-averaged equations for mass, momentum and energy
equation was described.   
This approach appears to be a robust physical description and a good
compromise between the full 3-D simulation and the necessity to
decrease the computational time.  
The model was satisfactorily applied to reproduce some analytical
solutions and to simulate a real lava flow event occurred during the  
1991-93 Etna eruption.
The good performance obtained in this preliminary version of the model
makes this approach a potential tool to forecast reliably lava flow
paths  to use for risk mitigation, although the used algorithm
should be improved for a better treatment of the source terms. 
\acknowledgement
This work was partially supported by the Gruppo Nazionale per la
Vulcanologia-INGV and the Italian Department of the Civil Protection.
This study was partially developed during the first author's PhD at
University of Bologna, Italy. 
%
%

\begin{thebibliography}{20}
\expandafter\ifx\csname natexlab\endcsname\relax\def\natexlab#1{#1}\fi

\bibitem[{{\it Alcrudo and Benkhaldoun\/}(2001)}]{alcben2001}
Alcrudo, F., and F.~Benkhaldoun, Exact solutions to the {Riemann} of the
  shallow water equations with a step, {\it Computers \& Fluids\/}, {\it 30\/},
  643--671, 2001.

\bibitem[{{\it Ambrosi\/}(1999)}]{amb99}
Ambrosi, D., Approximation of shallow water equations by {Riemann} solvers,
  {\it Int. J. for Numer. Meth. in Fluids\/}, {\it 20\/}, 157--168, 1999.

\bibitem[{{\it Barberi et~al.\/}(1993){\it Barberi, Carapezza, Valenza, and
  Villari\/}}]{barcar93}
Barberi, F., M.~Carapezza, M.~Valenza, and L.~Villari, The control of lava flow
  during the 1991-1992 eruption of {M}t. {E}tna, {\it J.\ Volcanol.\ Geotherm.\
  Res.\/}, {\it 56\/}, 1--34, 1993.

\bibitem[{{\it Burguete et~al.\/}(2002){\it Burguete, {Garcia-Navarro}, and
  Aliod\/}}]{burgar2002}
Burguete, J., P.~{Garcia-Navarro}, and R.~Aliod, Numerical simulation of runoff
  from extreme rainfall events in a mountain water catchment, {\it Nat.\ Haz.\
  Earth Syst.\ Sci.\/}, {\it 2\/}, 109--117, 2002.

\bibitem[{{\it Calvari et~al.\/}(1994){\it Calvari, Coltelli, Neri, Pompilio,
  and Scribano\/}}]{calcol94}
Calvari, S., M.~Coltelli, M.~Neri, M.~Pompilio, and V.~Scribano, The 1991-93
  {E}tna eruption: chronology and flow-field evolution, {\it Acta Vulcanol.\/},
  pp. 1--15, 1994.

\bibitem[{{\it Chinnayya et~al.\/}(2004){\it Chinnayya, {L}e{R}oux, and
  Seguin\/}}]{chiler2004}
Chinnayya, A., A.~{L}e{R}oux, and N.~Seguin, A well-balanced numerical scheme
  for the approximation of the shallow-water equations with topography: the
  resonance phenomenon, {\it International Journal on Finite Volumes\/}, 2004.

\bibitem[{{\it Costa and Macedonio\/}(2002)}]{cosmac2002}
Costa, A., and G.~Macedonio, Nonlinear phenomena in fluids with
  temperature-dependent viscosity: an hysteresis model for magma flow in
  conduits, {\it Geophys.\ Res.\ Lett.\/}, {\it 29\/}, 2002.

\bibitem[{{\it Costa and Macedonio\/}(2003)}]{cosmac2003}
Costa, A., and G.~Macedonio, Viscous heating in fluids with
  temperature-dependent viscosity: implications for magma flows, {\it Nonlinear
  Proc. Geophys.\/}, {\it 10\/}, 545--555, 2003.

\bibitem[{{\it Crisp and Baloga\/}(1990)}]{cribal90}
Crisp, J., and S.~Baloga, A model for lava flows with two thermal components,
  {\it J.\ Geophys.\ Res.\/}, {\it 95\/}, 1255--1270, 1990.

\bibitem[{{\it Ferrari and Saleri\/}(2004)}]{fersal2004}
Ferrari, S., and F.~Saleri, A new two-dimensional shallow water model including
  pressure effects and slow varying bottom topography, {\it ESAIM: Mathematical
  Modelling and Numerical Analisys\/}, {\it 38\/}, 211--234, 2004.

\bibitem[{{\it George\/}(2004)}]{geo2004}
George, D., {N}umerical {A}pproximation of the {N}onlinear {S}hallow {W}ater
  {E}quations with {T}opography and {Dry} {B}eds: {A} {G}odunov-{T}ype
  {S}cheme, Master's thesis, University of Washington, 2004.

\bibitem[{{\it Gerbeau and Perthame\/}(2001)}]{gerper2001}
Gerbeau, J., and B.~Perthame, Derivation of viscous {Saint}-{Venant} system for
  laminar shallow water; numerical validation, {\it Discret. Contin. Dyn.-B\/},
  {\it 1\/}, 89--102, 2001.

\bibitem[{{\it Heinrich et~al.\/}(2001){\it Heinrich, Piatanesi, and
  H\'ebert\/}}]{heipia2001}
Heinrich, P., A.~Piatanesi, and H.~H\'ebert, Numerical modelling of tsunami
  generation and propagation from submarine slumps: the 1998 {Papua} {New}
  {Guinea} event, {\it Geophys. J. Int.\/}, {\it 145\/}, 97--111, 2001.

\bibitem[{{\it Keszthely and Self\/}(1998)}]{kessel98}
Keszthely, L., and S.~Self, Some physical requirements for the emplacement of
  long basaltic lava flows, {\it J.\ Geophys.\ Res.\/}, {\it 103\/},
  27,447--27,464, 1998.

\bibitem[{{\it LeVeque\/}(1998)}]{lev98}
LeVeque, R., Balancing source terms and flux gradients in high-resolution
  {Godunov} methods: the quasi-steady wave-propagation algorithm, {\it J.
  Comput. Phys.\/}, {\it 146\/}, 346--365, 1998.

\bibitem[{{\it LeVeque\/}(2002)}]{lev2002}
LeVeque, R., {\it Finite Volume Methods for Hyperbolic Problems\/}, Cambridge
  University Press, 2002.

\bibitem[{{\it Monthe et~al.\/}(1999){\it Monthe, Benkhaldoun, and
  Elmahi\/}}]{monben99}
Monthe, L., F.~Benkhaldoun, and I.~Elmahi, Positivity preserving finite volume
  {Roe} schemes for transport-diffusion equations, {\it Comput. Methods Appl.
  Mech. Engrg.\/}, {\it 178\/}, 215--232, 1999.

\bibitem[{{\it Neri\/}(1998)}]{ner98}
Neri, A., A local heat transfer analysis of lava cooling in the atmosphere:
  application to thermal diffusion-dominated lava flows, {\it J.\ Volcanol.\
  Geotherm.\ Res.\/}, {\it 81\/}, 215--243, 1998.

\bibitem[{{\it Pieri and Baloga\/}(1986)}]{piebal86}
Pieri, D., and S.~Baloga, Eruption rate, area, and length relationships for
  some {H}awaiian lava flows, {\it J.\ Volcanol.\ Geotherm.\ Res.\/}, {\it
  30\/}, 29--45, 1986.

\bibitem[{{\it Shah and Pearson\/}(1974)}]{shapea74}
Shah, Y., and J.~Pearson, Stability of non-isothermal flow in channels - {III.}
  {T}emperature-dependent pawer-law fluids with heat generation, {\it Chem.
  Engng. Sci.\/}, {\it 29\/}, 1485--1493, 1974.

\end{thebibliography}
%

%
\end{article}
\end{document}